\documentstyle[multicol,amssymb,epsfig,prb,aps]{revtex}
\begin{document}
\def\cc#1{\kern#1pt
    \special{ps:  @beginspecial
    newpath -#1 2 div #1 2 div #1 2 div 0 360 arc fill
    @endspecial}}

\draft
\title{Hall effect in underdoped GdBa$_2$Cu$_3$O$_{7-\delta}$ thin films: evidence for
a crossover line in the pseudogap regime.}
\author{D. Matthey, S. Gariglio, B. Giovannini, and J.-M. Triscone}
\address{Condensed Matter Physics Departement, University of Geneva, Switzerland.}
\date{\today}
\maketitle
\begin{abstract}
 We report on measurements of the resistivity and Hall coefficient in underdoped
 GdBa$_2$Cu$_3$O$_{7-\delta}$ epitaxial thin films grown by off-axis magnetron sputtering. The films
 have been lithographically patterned allowing precise measurements of the
 temperature dependencies
 of the inverse Hall constant $R_H^{-1}$ and of the Hall angle $\theta_H$. We find
 that $R_H^{-1}$ is linear in temperature between $300\:K$ and the pseudogap temperature
 $T^*$, whereas $\cot(\theta_H)$
 displays a perfect $T^{2}$ temperature dependence between typically $300$ and $100\:K$.
 We observe
 for all the samples that the temperature at which the temperature dependence of
 $\cot(\theta_H)$
 deviates from the $T^{2}$ behavior is correlated to the temperature at which
 $R_H$ displays a peak.
 This characteristic temperature, found to lie between $T_c$ and $T^*$, does not
 depend markedly on the doping level and defines a new crossover line in the temperature
 versus doping phase diagram. We tentatively relate these findings to recent
 high frequency conductivity and Nernst effect experimental
results, and we briefly discuss the possible
 consequences for competing theories for the pseudogap state of the cuprates.
\end{abstract}

\begin{multicols}{2}
\section{Introduction}
Among the anomalous normal state transport properties observed in
high temperature superconductors, the temperature dependence of
the Hall constant is probably one of the most striking: it
displays a behavior which is totally different from the one
observed in simple metals. Indeed, it is generally observed, in
particular in REBa$_2$Cu$_3$O$_{7-\delta}$ (RE rare earth)
compounds, that the temperature dependence of the Hall
coefficient, $R_H(T)$, goes as $1/T$ over a wide temperature range
and presents an ill-understood peak above $T_c$, whereas the
cotangent of the Hall angle ($\cot(\theta_H) =
\rho_{xx}/\rho_{xy}$ where $\rho_{xx}$ and $\rho_{xy}$ are
respectively the longitudinal and Hall resistivity) varies as
$T^{2}$ over a wide temperature range. One way to analyze the
temperature dependence of the Hall effect is to consider two
different relaxation rates. Experimental reports
\cite{carringtonprl92}, showing that the cotangent of the Hall
angle obeys a $T^{2}$ law even if $R_H^{-1}(T)$ or $\rho_{xx}(T)$
are strongly nonlinear suggest that the cotangent of the Hall
angle is a function of a single relaxation rate, $\tau_H^{-1}$,
whereas the resistivity $\rho_{xx}$ would be related to another
relaxation rate $\tau_{tr}^{-1}$. Other scenarios based on a
temperature dependent carrier
concentration\cite{luocondmat00,hirschphc92} or on a scattering
time that depends on the Fermi surface
location\cite{stojkovicprb97,ioffeprb98,caoprb00} have been
proposed. When considering two relaxation rates, the Hall constant
$R_H$ depends on both $\tau_H$ and $\tau_{tr}$: $R_H^{-1}=
\cot(\theta_H)\cdot H/\rho_{xx}\sim\tau_H^{-1}/\tau_{tr}^{-1}$.
The $T^{2}$ temperature dependence of the cotangent of the Hall
angle, $\cot(\theta_H)=AT^{2}+C$, has been theoretically predicted
\cite{andersonprl91,levinprb92,stojkovicprl96,caoprb00}, and
observed experimentally in many systems \cite{chienprl91}. In the
approach proposed by Anderson \cite{andersonprl91,andersonbook}
the two scattering rates reflect spin and charge separation with
the Hall angle being directly related to the spinon relaxation
rate. In this view, magnetic impurities bring another scattering
channel for the spinons and thus directly affect the behavior of
the Hall angle as first observed experimentally by T. R. Chien, Z.
Z. Wang and N. P. Ong \cite{chienprl91}.

Recent experimental results\cite{timuskrpp99,com10} and
theoretical ideas on the nature of the pseudogap
\cite{emerynat95,devillardprl00,emeryprl00,vojtacondmat00,chakravartycondmat00,uemuracondmat00}
and its influence on Hall measurements
\cite{abeprb99,konstantinovicprb00}, the possible presence of
charge inhomogeneities \cite{nodasci95,mooknat00}, or the presence
of vortex like excitations in the pseudogap phase
\cite{corsonnat99,xunat00} have changed our view of the normal
state. In this paper, we revisit the question of the Hall effect
by measuring the Hall constant and the Hall angle in the pseudogap
phase of epitaxial underdoped GdBa$_2$Cu$_3$O$_{7-\delta}$ films.
The results are presented and discussed in the context of the
recent findings mentioned above.

\section{Samples and experiment}
A series of GdBa$_2$Cu$_3$O$_{7-\delta}$ thin films were grown by off-axis
magnetron sputtering on (100) SrTiO$_3$ substrates. The temperature of the substrates was typically $700\:^{\circ}C$, and the $Ar + O_2$
sputtering pressure $0.15\:Torr$
  $(O_2/Ar=0.4)$. \cite{trisconerpp97} The oxygen content in the films was changed
  by varying the oxygen pressure during the cooling procedure \cite{com1}. The structural
  crystalline quality of the films was checked by x-ray
  diffraction analysis. $\theta-2\theta$ and phi scans indicate an epitaxial growth with
   the $c$ axis oriented perpendicular to the substrate surface. Low angle x-ray reflectivity
   oscillations as well as finite size effects around
   the 001 peaks allow determination of the individual
   sample thickness, which was varied from $20$
   to $60\:nm$ in this study. To prevent degradation during the lithographic processes, the
   films were protected by the subsequent in-situ deposition of an amorphous insulating layer.
   All the samples were photolithographically patterned using ion
   milling with a $100\:\mu m$ bridge wide. The relatively low thickness of the films allowed
   us to increase the measured voltages and to reduce the noise of the measurements. The $dc$ Hall
   voltage was measured using a three voltage contact technique and by adjusting the Hall
   voltage to zero in zero magnetic field. At each temperature, the Hall voltage
   was measured by sweeping the magnetic field between $-6$ and $+6$ Tesla. During
   each measurement, the temperature, measured with a cernox resistance, whose
   magnetoresistance is negligibly small, was stable within $\pm$ $5\:mK$.

\section{Results and discussion}
\begin{figure}
\centering
\includegraphics[scale=0.8]{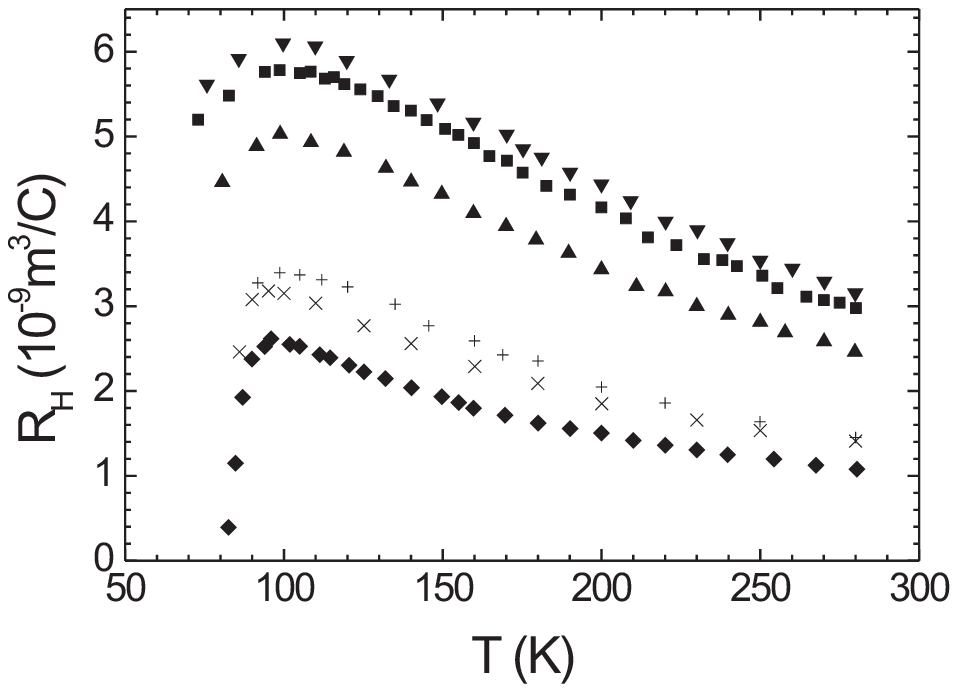}
\caption{Temperature dependence of the Hall constant $R_H$ for a
series of GdBa$_2$Cu$_3$O$_{7-\delta}$ thin films. The curves
correspond to films with critical temperatures of $84.6\:K$
$(\blacklozenge)$, $80.9\:K$ (X), $79.0\:K$ $(+)$, $62.7\:K$
$(\blacktriangle)$, $48.9\:K$ $(\blacksquare)$ and $53.1\:K$
$(\blacktriangledown)$.} \label{figure1}
\end{figure}

Figure \ref{figure1} shows the Hall constant $R_H$, as a function
of temperature, for a series of GdBa$_2$Cu$_3$O$_{7-\delta}$ thin
films. On Fig. \ref{figure1} the curves correspond to films with
critical temperatures of $84.6\:K$, $80.9\:K$, $79.0\:K$,
$62.7\:K$, $48.9\:K$ and $53.1\:K$. As can be seen, $T_c$
[Ref.\onlinecite{com2}] scales with the inverse Hall coefficient
$R_H^{-1}$. Since $R_H^{-1}$ depends on the temperature, the
relation between the doping level and $R_H^{-1}$ is not
straightforward. Here, we will simply use, as others
\cite{jinprb98}, the value of $R_H^{-1}$ at 100 $K$ as a parameter
related to the doping level. We find that the $T_c$ versus
$R_H^{-1}(100\:K)$ curve obtained is in good agreement with
YBa$_2$Cu$_3$O$_{7-\delta}$ single crystals data\cite{itoprl93}.
This observation is consistent with a $T_c$ reduction and a change
in doping in these thin films essentially related to a change in
oxygen content. In Fig. \ref{figure1} the highest $T_c$ sample has
a resistivity which is typically twice the value found in
optimally doped thin films \cite{carringtonprb93}. The value of
$R_H^{-1}(100\:K)$ for this same sample is slightly less than half
the value found in optimally doped thin films
\cite{carringtonprb93}, implying that, in this thickness range,
our films are not fully oxidized \cite{com3}. One can also notice
on Fig. \ref{figure1} that the maximum in $R_H$ occurs in the
normal state around $100\:K$ for each sample and does not seem to
be related to the $T_c$ of the sample. This point will be
discussed below in detail.

\begin{figure}
\centering
\includegraphics[scale=1]{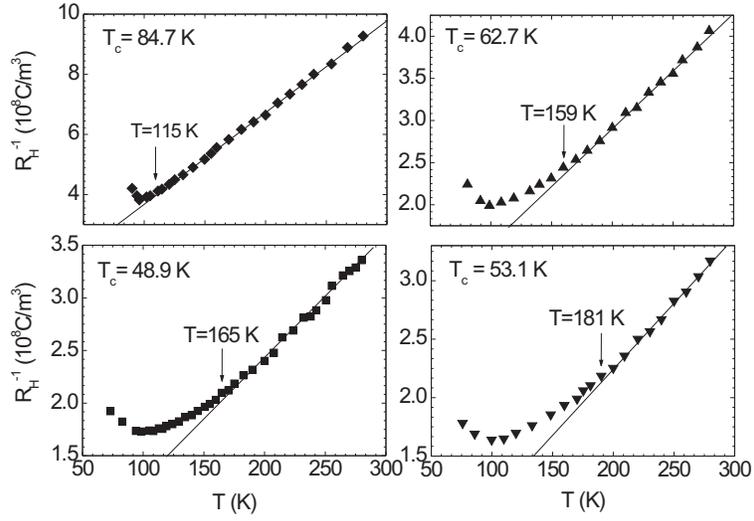}
\caption{Temperature dependence of $R_H^{-1}$ for four
representative samples. The temperature at which the inverse Hall
constant deviates from the linearity is shown by an arrow for each
sample. This characteristic temperature can be associated with the
pseudogap temperature $T^{*}$.} \label{figure2}
\end{figure}

In Fig. \ref{figure2} the temperature dependence of $R_H^{-1}$ is
shown for four representative samples. For the almost optimally
doped sample with $T_c=84.7\:K$, $R_H^{-1}$ is linear in
temperature over a wide temperature range as observed
experimentally in optimally doped YBa$_2$Cu$_3$O$_{7-\delta}$
crystals \cite{ongbook}. The temperature at which the inverse Hall
constant deviates from linearity is shown by an arrow for each
sample. For the highest $T_c$ sample the deviation in linearity is
found to be very close to the peak in $R_H$ (minimum of
$R_H^{-1}$). For more underdoped samples, the temperature at which
the deviation occurs is shifted, the lower the $T_c$, the higher
this temperature. These data are found to be consistent with the
results of R. Jin and H. R. Ott \cite{jinprb98} who made a
systematic study of the Hall effect on YBa$_2$Cu$_3$O$_{7-\delta}$
single crystals. They found that the temperature at which
$R_H^{-1}$ deviates from linearity is in good agreement with the
temperature at which anomalies in the temperature dependence of
other physical properties are reported, these anomalies being
associated with the opening of the pseudogap at the temperature
$T^*$.\cite{com8}

\begin{figure}
\centering
\includegraphics[scale=1]{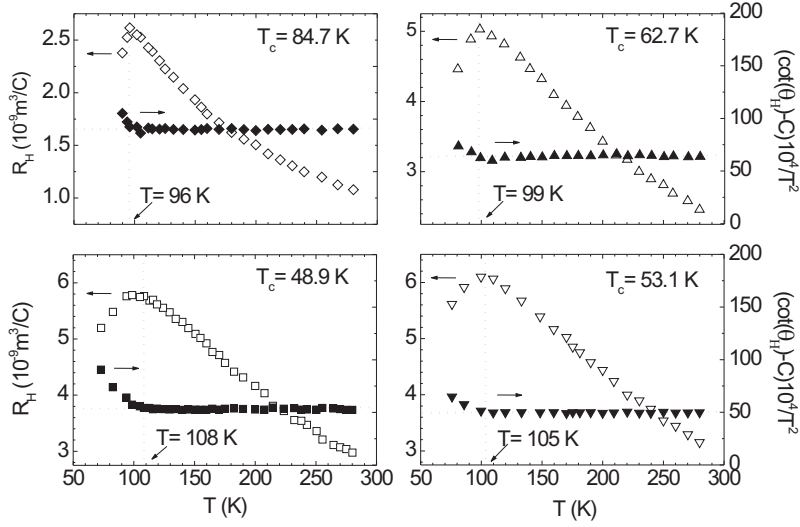}
\caption{$(\cot(\theta_H)-C)/T^2$ (right axis) and $R_H$ (left
axis) for four selected samples are shown as a function of
temperature. $(\cot(\theta_H)-C)/T^2$ is temperature independent
for all the samples until a characteristic temperature $T^{'}$
which is well correlated with the temperature at which $R_H$
displays a peak.} \label{figure3}
\end{figure}

We turn now to the analysis of the behavior of $\cot(\theta_H)$.
Figure \ref{figure3} shows for four selected samples
$(\cot(\theta_H)-C)/T^{2}$ and $R_H$ as a function of temperature.
Our data on the temperature dependence of $\cot(\theta_H)$ show,
over a temperature range extending from $100\:K$ to $300\:K$, a
perfect $T^2$ dependence for all the doping levels investigated.
No anomalies at temperatures close to $T^*$ (or at the temperature
at which $R_H^{-1}(T)$ deviates from linearity) are found in
$\cot(\theta_H)(T)$. As shown in Fig. \ref{figure3},
$(\cot(\theta_H)-C)/T^{2}$ is temperature independent for all the
samples until a characteristic temperature at which a deviation is
observed. We notice, as apparent on Fig. \ref{figure3}, that this
temperature is essentially not affected by the sensitivity
criterium used to define the deviation in the $T^2$ temperature
dependence of $\cot(\theta_H)$. As can be seen also in Fig.
\ref{figure3}, this second characteristic temperature $T^{'}$, is
well correlated with the temperature at which $R_H$ displays a
peak and is almost independent of the doping level in the range
investigated. At this temperature $T^{'}$ no anomaly is observed
in the resistivity, the peak in
$R_H(\sim\rho_{xx}/\cot(\theta_H))$ has to be attributed to the
change in $\cot(\theta_H)(T)$.

Since no deviation in $\cot(\theta_H)$ is observed at high
temperature (close to $T^*$), we conclude in the two relaxation
times picture, that the deviation in $R_H^{-1}=
\cot(\theta_H)\cdot H/\rho_{xx}\sim\tau_H^{-1}/\tau_{tr}^{-1}$ has
to be attributed to a change in $\tau_{tr}^{-1}$, a diminution of
the scattering rate related to the opening of the pseudogap. Since
$\rho_{xx}\sim\tau_{tr}^{-1}$, a deviation from the linear
temperature dependence of the resistivity should be observed at
the same characteristic temperature. A detailed analysis of the
resistivity is difficult however, since, as often observed in
underdoped thin films, the resistivity is not linear with
temperature in this temperature range. It should be noticed that
the $T^{2}$ temperature dependence of $\cot(\theta_H)$ is observed
up to temperatures much higher than $T^{*}$ implying that the
mechanism at the origin of the $T^2$ behavior extends at
temperatures above the pseudogap phase.

\begin{figure}
\centering
\includegraphics[scale=0.8]{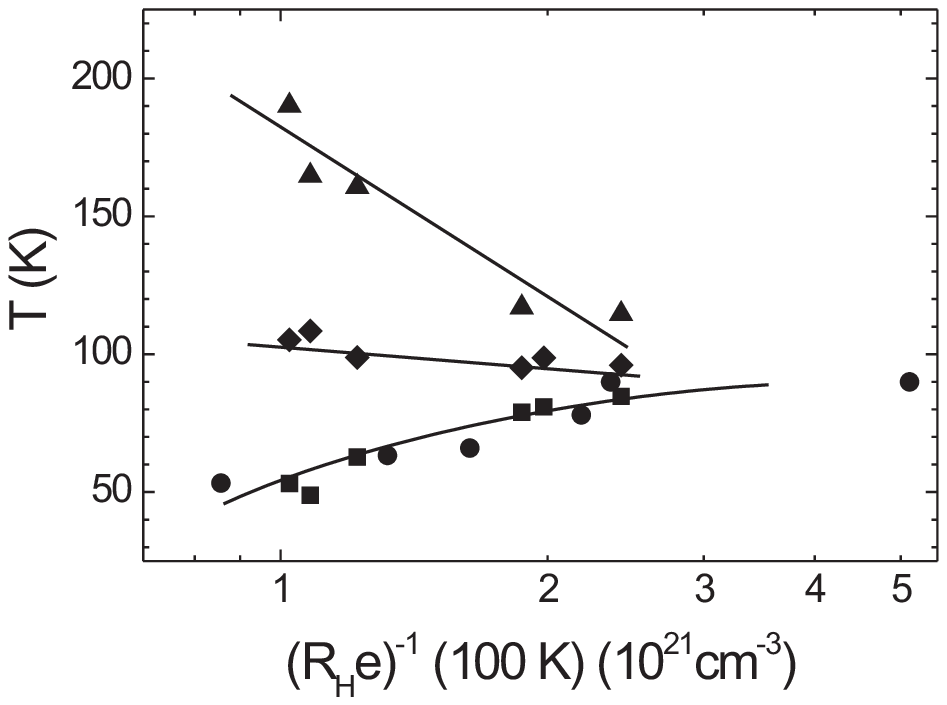}
\caption{The phase diagram $T$ versus $R_H^{-1}(100\:K)$ (logscale) summarizes the
results of the paper. $T_c$ versus $R_H^{-1}(100\:K)$
$(\blacksquare)$,
along with single crystals data from Ito \textit{et al}\cite{itoprl93} (\cc{5}),
the temperature at which $R_H^{-1}(T)$ deviates from linearity associated to the
pseudogap temperature $T^*$ $(\blacktriangle)$ and $T^{'}$ $(\blacklozenge)$, the
temperature at which a deviation in the $T^2$ behavior of
$\cot(\theta_H)$ is observed, are plotted.}
\label{figure4}
\end{figure}

In Fig. \ref{figure4}, we present a phase diagram $T$ versus
$R_H^{-1}(100\:K)$ that summarizes the results of the paper. First
$T_c$ versus $R_H^{-1}(100\:K)$ is plotted (squares) along with
single crystals data\cite{itoprl93} (circles). The line is a guide
to the eyes. As can be seen the critical temperatures of films and
crystals fall on the same curve. The temperatures at which a
deviation in the linear temperature dependence of $R_H^{-1}(T)$ is
observed are indicated by triangles. These points follow the
temperature dependence of the pseudogap temperature $T^*$ as
discussed above. Finally, the diamonds are the well defined
temperatures $T^{'}$ at which a deviation in the $T^2$ behavior of
$\cot(\theta_H)$ occurs (or at which $R_H$ is maximum). $T^{'}$
falls between $T_c$ and $T^*$, clearly separating the pseudogap
phase in two regions.

Before discussing this phase diagram and in particular the
temperature $T^{'}$, one can compare these results to very recent
experiments on Bi$_2$Sr$_2$Ca$_{n-1}$Cu$_n$O$_y$ films
\cite{konstantinovicprb00} which focalize on the doping dependence
of the Hall response. In this system Z. Konstantinovi\'c, Z. Z. Li
and H. Raffy do not observe a correlation between the peak in
$R_H$ and the deviation in $\cot(\theta_H)$. Their data, however,
show that the temperature at which a deviation in
$\cot(\theta_H)(T)$ occurs is much lower than $T^*$, in agreement
with the data presented here. We can also relate our data to other
recent experimental findings. Analysis of the temperature
dependence of $R_H(T)$ have suggested that the peak in $R_H(T)$ is
related to the pseudogap \cite{itoprl93,abeprb99}. It is however
clear from the experimental data shown in Fig. \ref{figure4}, that
the temperature of the maximum in $R_H$ is well defined and does
not follow the pseudogap temperature dependence, suggesting that
the latter is not, at least directly, responsible for the behavior
observed.

There have also been reports on Bi$_2$Sr$_2$CaCu$_2$O$_{8+\delta}$
and La$_{2-x}$Sr$_x$CuO$_4$ compounds of the presence of vortex
like excitations in the pseudogap phase. High frequency
conductivity experiments \cite{corsonnat99} and measurements of
the Nernst effect \cite{xunat00} have shown that vortex like
excitations are found at temperatures up to $100-150\:K$. Although
measured on different systems, these temperatures are above $T_c$
but below the pseudogap temperature $T^*$, as the characteristic
temperatures $T^{'}$ measured in this work. Since experimentally
\cite{com5} without implying a particular scenario, the Hall angle
has been shown to be sensitive to magnetic impurities through
$\tau_H^{-1}$, an interesting possibility is that vortex like
excitations may affect the Hall angle. These vortex like
excitations could be seen as a parallel, temperature dependent
channel for magnetic scattering modifying the Hall angle and
indirectly $R_H$. The enhancement of $\cot(\theta_H)$ due to this
additional scattering source would be responsible, in this system
for the break in the temperature dependence of $R_H$.

The Nernst effect experiments, the high frequency experiments and
the Hall data shown here point to a new crossover in the phase
diagram. From an experimental point of view, three crossover
temperatures are thus observed in underdoped cuprates: from weak
($T_0$) to strong pseudogap at $T^*$ and, at lower temperature,
the "new" crossover line discussed above. The crossover from weak
to strong pseudogap has been intensively discussed in the
literature theoretically and experimentally\cite{timuskrpp99}.
Several theoretical approaches have been suggesting or calculating
the existence of an additional structure in the strong pseudogap
phase\cite{emeryprl00,vojtacondmat00}. In particular, phase
diagrams with a new temperature $T_{charge}$ at which charge
ordering occurs\cite{mooknat00} have been discussed in the
presence of charge inhomogeneities or
stripes\cite{chakravartycondmat00}. Experimentally, T. Noda, H.
Eisaki and S. Uchida have measured the Hall effect in
La$_{2-x-y}$Nd$_y$Sr$_x$CuO$_4$ [Ref. \onlinecite{com9}] and have
related the sharp decrease in $R_H$ to one dimensional charge
transport \cite{nodasci95}. We note also that recently, P.
Devillard and J. Ranninger have predicted that there should exist
a new characteristic temperature $T^*_B$ in the pseudogap phase,
between $T^*$ and $T_c$, that corresponds to the temperature at
which the electron pairs, formed at higher temperature ($T^*$),
become itinerant, defining an upper limit for partial Meissner
screening \cite{devillardprl00} (and vortex like excitations).

There are perhaps two main tentative theoretical frameworks for
the pseudogap regime. In one, the pseudogap state is fluctuating
superconductivity: the system is superconducting over short
distances and time scales, but phase fluctuations destroy the long
range superconducting order. The second main theoretical framework
is related to the proximity of these systems to Mott insulators,
and attempts to describe these strongly correlated systems with
various schemes of spin charge separation or gauge
theories\cite{leephc99,senthilcondmat99}. The Hall effect
experiments presented in this paper raise in fact two related
questions. Firstly whether the difference between the Hall and the
resistivity temperature dependencies entails the existence of two
different relaxation mechanisms and times, or whether it should be
explained in a different way. If two different scattering times do
indeed characterize the Hall effect situation in high temperature
superconductors, then one must somehow reconcile this fact with
the existence of the lower crossover. Unfortunately, the various
spin charge separation theories, that provide in principle a
natural explanation for the different relaxation
times\cite{andersonprl91,andersonbook}, have up to now, to our
knowledge, not predicted a second crossover in the strong
pseudogap regime. This difficulty could possibly be resolved if
one assumes that the holons or chargons behave as 2D bosons: the
new temperature $T^{'}$ could then be the critical temperature of
the corresponding 3D system as recently discussed by Y. J.
Uemura\cite{uemuracondmat00} in the context of "preformed pairs".
On the other hand, if the lower crossover is connected with the
loss of short distance phase coherence between "preformed pairs",
and therefore the disappearance of vortices, one is left in this
phase fluctuation picture with the problem of finding a mechanism
for the existence of two different relaxation times, and this, way
above $T^{*}$.

\section{Conclusion}
In conclusion, we have studied the transport properties,
resistivity and Hall effect, in a series of underdoped
GdBa$_2$Cu$_3$O$_{7-\delta}$ epitaxial thin films, grown by
off-axis magnetron sputtering. We find that the inverse Hall
constant is linear over a temperature range that decreases as
$T_c$ decreases. The temperature at which a deviation in
$R_H^{-1}$ occurs is close to the pseudogap temperature $T^*$
measured in this material. We also find that $\cot(\theta_H)$
displays a perfect $T^2$ temperature dependence over a large
temperature range extending from about $100\:K$ to $300\:K$. We
observe for all the samples that the temperature at which the
temperature dependence of the Hall angle deviates from the $T^2$
behavior is precisely correlated to the temperature at which $R_H$
displays a peak. This characteristic temperature, found to lie
between $T_c$ and the pseudogap temperature $T^*$, does not depend
markedly on the doping level and defines a new crossover in the
phase diagram.
\newline

We thank Z. Te$\rm{\breve{s}}$anovi\'c for bringing to our
attention the possible importance of vortices in the pseudogap
phase, D. Marr\'e for help in the initial stage of this project
and D. Chablaix for technical help. This work was supported by
Swiss National Science Foundation.


\end{multicols}

%
%

%
%

\begin{thebibliography}{10}
\bibitem{com1}
Although for similar cooling procedures{,} the $T_c$ of samples
with different
  thicknesses were different{,} we found that the $T_c$ versus $R_H^{-1}(100K)$
  data points fall on the same curve{,} irrespective of the film thickness.
\bibitem{com2}
$T_c$ is defined as the temperature at which the resistivity{,} on
a
  logarithmic scale{,} falls in the noise level of the measurements.
\bibitem{com3}
The difficulty to fully oxidize these samples might be related to
strains in
  very thin films.
\bibitem{com8}
The influence of the pseudogap on the Hall constant in
YBa$_2$Cu$_4$O$_8$ and
  in La$_{2-x}$Sr$_x$CuO$_4$ is respectively discussed in Ref.
  \onlinecite{bucherprl93} and Ref. \onlinecite{hwangprl94}.
\bibitem{com9}
This particular compound has a structural phase transition at a
temperature
  $T_0$ and the low temperature phase stabilizes charges inhomogeneities
  (stripes).
\end{thebibliography}
\end{document}